\documentclass[twocolumn,showpacs,preprintnumbers,amsmath,amssymb]{revtex4}

\usepackage{graphicx}
\usepackage{dcolumn}
\usepackage{bm}

\begin{document}


\title{A  Double Layer  Electromagnetic Cloak And GL EM Modeling}
\author{Ganquan Xie}
 \altaffiliation[Also at ]{GL Geophysical Laboratory, USA, glhua@glgeo.com}
\author{Jianhua Li, Feng Xie, Lee Xie}%
 \email{GLGANQUAN@GLGEO.COM}
\affiliation{%
GL Geophysical Laboratory, USA
}%
\hfill\break
\date{\today}
\begin{abstract}
In this paper, we propose a novel electromagnetic (EM) cloaking structure
that consists of two annular layers between three spherical shells and model its
performance numerically and theoretically by using a Global and Local EM field (GL) method. 
The two annular layers contain distinct cloaking
materials: the outer layer provides invisibility; the inner layer is fully
absorbing. The cloaking materials are weakly degenerative.
The wavefield from an EM source located outside the cloak propagates as in
free space outside the outer shell, never be disturbed by the cloak and 
does not penetrate into the inner absorbing layer and concealment.
 The field of a source located inside inner layer or the cloaked 
concealment is completely absorbed by the inner layer and never 
reaches the outside of the middle shell, Moreover, the EM wavefield excited 
in concealment is not disrupted by the cloak. 
There exists no Maxwell EM wavefield can be excited in a single layer
cloaked concealment which is filled by normal material.  Moreover,
we find a negative dielectric and positive susceptibility
metamaterial to fill into the concealment, such that 
the  interior EM wave propagates from the concealment to free space through the
single layer cloak. Therefore, the double layer cloak is important
for complete invisibility.
Numerical simulations and
theoretical analysis verifying 
these properties are performed using the GL EM 
modeling method that we described in this paper.
\end{abstract}


\maketitle

\section{\label{sec:level1}INTRODUCTION} 
We propose a novel kind of electromagnetic (EM) double layer cloaking structure 
(a ``Double Layer EM cloak") that consists of two annular layers between three 
spherical shells, $R_1\le r \le R_2$ and $R_2 \le  r \le R_3$. 
Distinct metamaterials are situated in the two layers: the material in the outer layer has properties that
provide invisibility, while material in the inner layer is fully
absorbing, which can also be useful for making a complete absorption boundary condition. 
We analyze the performance of the double layer cloak using a
Global and Local EM field modeling ``GL method" in time domain that we 
developed in this paper, its versions in frequency domain have been developed in 
our other papers [1-3]. 
Pendry et al. [4] proposed a single layer EM cloak material by using
a coordinate transformation and ray tracing in 2006. In their cloak, the ray being 
bending and re-direction around central sphere object and cannot penetrate into it.  
The cloak device like the vacuum and does not disturb exterior wave field.  
Later papers studied 2-D plane wave propagation through a cloaking layer are published in Physical
review etc. Journals; Mie scattering analytical method for the sphere cloak is studied in [5]; 
Numerical methods, including finite-difference time-domain method in [6]
and finite-element method in [7]  are presented in which many relative research 
work papers are cited. Most researchers paid attention for exterior EM wave field propagation through the cloak. Two authors have
studied the cloak's effect on sources located inside concealed region
[8] and [9]. The authors of [8] stated that "when these conditions are over-determined, finite energy solutions typically do not exist." 
They wrote that the single layer cloak is insufficient and proposed a double coating to solve this problem. However, 
there also is disputing for this problem.  
In many numerical simulations using the GL method, 
we discovered and verified cases where a field satisfying Maxwell's equations cannot be excited by a local
source inside of the single layer cloaked concealment which is filled with normal materials. 
For the above case of EM field excited inside the concealment, our
simulations are either divergent or become chaotic when the field propagates
to the inner boundary of the single cloaking layer. These simulations remind
us to prove that there exists no Maxwell EM wavefield can be excited by source
inside of the single layer cloaked concealment which is filled by normal material.¡¨
The theoretical proof of the statement is proved in this paper by using  he GL method.
Moreover,
we find a negative dielectric and positive susceptibility
metamaterial to fill into the concealment, such that 
the  interior EM wave propagates from the concealment to free space through the
single layer cloak. Therefore, the double layer cloak is important
for complete invisibility.

The double layer cloak proposed here overcomes these difficulties described above by situating
a second, absorbing layer inside the outer cloaking layer. 
The wavefield excited from an EM source located outside the cloak propagates as in
free space outside the outer shell and never be disturbed by the cloak;
it propagates around and through the outer layer 
cloak material and does not penetrate into the inner layer and innermost concealment.
The field of a source located inside inner layer or the cloaked 
concealment is completely absorbed by the inner layer and never 
reaches the outside of the middle shell. 
Un Maxwaell physical behavior at the
inner boundary of single cloaking layer does not appear in our simulations of the double layer cloak. The EM wavefield excited 
in concealment is not disrupted and has no reflection by the double layer cloak. In which the EM enviroment in the concealment is keeped
to be normal and has no be changed.  As a result,
a double layer EM cloak appears to be more thoroughly concealing and more
robust than single layer structures.
Moreover, the metamaterials situated in the double layer cloak are
weakly degenerative.

The analytical method and numerical method for physical simulation have 
been developed separately in history. The GL method consistently 
combines both analytical and numerical approaches.  The GL method 
does not need to solve large matrix equations, it only needs to solve
 $3 \times 3$ and $6 \times 6$ matrix equations.  Moreover, the GL 
method does not prescribe any artificial boundary, and does not need 
a PML absorption condition to truncate the infinite domain.  The Finite 
Element Method (FEM) and Finite Difference Method (FDM) have 
numerical dispersions which confuse and contaminate the physical 
dispersion in the dispersive metamaterials. The frequency 
limitation is also a difficulty of FEM and FDM. The ray tracing is used for wave 
direction only that is not suitable to study full wave field propagation.

The GL method is a significant scattering process which reduces the 
numerical dispersion and is suitable to simulate physical wavefield 
scattering in the materials, in particular, for dispersive materials. Born 
approximation is a conventional method in the quantum mechanics and 
solid state physics. However, it only has one iteration in the whole 
domain which may not be accurate in the high frequency and high 
contrast materials. The GL method divides the domain as a set of 
sub-domains or sub-lattices which are as small as
accurate requirement.  The Global field is updated by the local field 
from the interaction between the global field and local sub-domain 
materials successively. Once all sub-domain materials are scattered,
 the GL field solution obtained turn out to be more accurate than the 
Born approximation. Moreover, the GL method can be mesh-less, including arbitrary 
geometry sub-domains, such as rectangle, cylindrical and spherical 
coordinate mixed coupled together.  It is full parallel algorithm.  
The GL method advantages help overcome many historical obstacles 
described in detail in [1].  The theoretical foundation of the GL method 
is described in the paper [2]. In particular, for the radial dependent cloak
materials, the GL method can be reduced to the system of the one dimensional
GL processes in radial interval $[R_1,R_3]$ by using spherical harmonics expansion. 
We have used the GL modeling [1-2]
and inversion technique in [3] to simulate many radiaul cloak metamaterials, 
nanometer materials, periodic photonic crystals etc.
These simulations show that the 
GL method is fast and accurate. 

 The introduction is described 
in Section 1.
The rest of this paper is organized as follows: 
Section 2 describes the geometry and material properties of the proposed double layer cloak. Section 3 
contains a brief description of the EM integral equations and their solution by the GL method. 
These equations are used in Section 4 to prove certain theorems about double layer and single layer cloaks. 
Section 5 shows numerical simulations of the double layer cloak described in Section 2. The advantages
of The GL cloak and GL modeling are presented in Section 6. Section 7 presents our conclusions.

\section{\label{sec:level1} Double Layer Electromagnetic Cloak}

In this section, by simulations using GL EM modeling  and inversion  in time domain which is presented in next section,
we propose a novel electromagnetic (EM) cloaking structure
that consists of two annular layers between three spherical shells.  An inner layer EM anisotropic cloak metamaterial is
situated in the inner annular layer; Outer layer EM cloak material is situated in outer layer.

\subsection {Inner Layer EM Anisotropic Cloak Material}
On the inner sphere annular layer domain, $\Omega _{I}  = \left\{ {r:R_1  \le r \le R_2 } \right\},$ 
we propose an anisotropic material as follows, 
\begin{equation}
\begin{array}{l}
 \left[ D \right]_{I}  = diag\left[ {\bar \varepsilon _i,\bar \mu _i } \right], \\ 
 \bar \varepsilon _i  = diag\left[ {\varepsilon _{r,i},\varepsilon _{\theta ,i} ,\varepsilon {}_{\phi ,i}} \right]\varepsilon _0,  \\ 
 \bar \mu _i  = diag\left[ {\mu _{r,i} ,\mu _{\theta ,i} ,\mu {}_{\phi ,i}} \right]\mu _0,  \\ 
\varepsilon _{r,i}  = \mu _{r,i}  
= \frac{{\left( {R_2  - r} \right)}}{{r^2 }}\left( {R_1  + \log \frac{{R_2  - R_1 }}{{R_2  - r}}} \right)^2 \\
\varepsilon _{\theta ,i}  = \varepsilon _{\phi ,i}  = \mu _{\theta ,i}  = \mu _{\phi ,i}  = \frac{1}{{R_2  - r}},\\
 \end{array}
\end{equation}
The $\Omega_{I}$ is called as the inner layer cloak, the material, $ \left[ D \right]_{I}  = diag\left[ {\bar \varepsilon _i ,\bar \mu _i } 
\right]$ in (1), is the anisotropic  inner layer cloak metamaterial.  The field of a source located inside inner layer or the cloaked 
concealment is completely absorbed by the inner layer and never 
reaches the outside of the middle shell, also, the EM wavefield excited 
in concealment is not disrupted by the cloak. The inner layer metamaterial, in equation (1),  cloaks outer space from the local field excited in the inner layer
and concealment,  which can also be useful for making a complete absorption boundary condition
to truncate infinite domain in numerical simulation. 

\subsection {Outer Layer Anisotropic Cloak Material}
Let the outer sphere annular domain $\Omega _{O}  = \left\{ {r:R_2  \le r \le R_3 } \right\}$ be the outer layer cloak
with the following anisotropic  outer layer cloak material,
\begin{equation}
\begin{array}{l}
 \left[ D \right]_{O}  = diag\left[ {\bar \varepsilon _o ,\bar \mu _o } \right], \\ 
 \bar \varepsilon _o  = diag\left[ {\varepsilon _{r,o} ,\varepsilon _{\theta ,o} ,\varepsilon {}_{\phi ,o}} \right]\varepsilon _0, \\ 
 \bar \mu _o  = diag\left[ {\mu _{r,o} ,\mu _{\theta ,o} ,\mu {}_{\phi ,o}} \right]\mu _0,  \\ 
  \varepsilon _{r,o}  = \mu _{r,o}  = \frac{{R_3 }}{r}\frac{{r^2  - R_2^2 }}{{r^2 }}\frac{{\sqrt {r^2  - R_2^2 } }}{{\sqrt {R_3^2  - R_2^2 } }}, \\ 
 \varepsilon _{\theta ,o}  = \mu _{\theta ,o}  = \varepsilon _{\phi ,o}  = \mu _{\phi ,o}  \\ 
  = \frac{{R_3 }}{{\sqrt {R_3^2  - R_2^2 } }}\frac{r}{{\sqrt {r^2  - R_2^2 } }}. \\ 
\end{array}
\end{equation}
Outer layer cloak provides invisibility, does not disturb exterior EM wave field, and cloaks the concealment from the exterior EM wavefield.

\subsection {GL Double Layer Cloak}
The  inner cloak $\Omega_{I}$ domain and outer cloak $\Omega_{O}$ domain 
are bordering on the sphere annular surface $r=R_2$.
We assemble the $\Omega_{I}$ as the inner sphere annular domain and $\Omega_{O}$ as
the outer sphere annular domain and make them coupling on their interface boundary annular surface  $r=R_2$
as follows,
\begin{equation}
\begin{array}{l}
 \Omega _{GL}  = \Omega _{I} \bigcup {\Omega _{O} }  \\ 
  = \left\{ {r:R_1  \le r \le R_2 } \right\}\bigcup {\left\{ {r:R_2  \le r \le R_3 } \right\}}  \\ 
  = \left\{ {r:R_1  \le r \le R_3 } \right\}, \\ 
 \end{array}
\end{equation}
and situate the Double Layer anisotropic dielectric and susceptibility tensors $\left[ D \right]_{GL}$ on the $\Omega _{GL}$  as follows,
\begin{equation}
\left[ D \right]_{GL}  = \left\{ {\begin{array}{*{20}c}
   {\left[ D \right]_{I} ,r \in \Omega _{I} }  \\
   {\left[ D \right]_{O} ,r \in \Omega _{O} }.  \\
\end{array}} \right.
\end{equation}
The double layer EM cloak materials $ \left[ D \right]_{I} = diag\left[ {\bar \varepsilon _i,\bar \mu _i } 
\right]$ in (1) on the $\Omega _{I}$  and  outer layer cloak material $ \left[ D \right]_{O}  = diag\left[ {\bar \varepsilon _o ,\bar \mu _o } 
\right]$ in (2) on $\Omega _{O}$  are assembled into the doubled layer cloak material $ \left[ D \right]_{GL}$  on
the domain $\Omega _{GL}$. The domain $\Omega _{GL}$ with the metamaterial $\left[ D \right]_{GL}$ in (4) 
is called as the  double layer cloak which is named GL cloak (GLC). GL cloak means that the outer layer has invisibility, never disturb exterior
wavefield and cloaks the Local concealment from the exterior Global field,  
while the inner layer cloaks outer Global space from interior wavefield excited in the Local concealment.
The properties of the double layer material are studied by GL method simulations and analysis in next sections.
\subsection {GLPS Double Layer Cloak}
To compare the properties between the GLO cloak and PS cloak, we use the GLI cloak,
in equation (1),  as inner layer and the PS cloak in [4] with the following anisotropic dielectric and susceptibility,
\begin{equation}
\begin{array}{l}
 \left[ D \right]_{PS}  = diag\left[ {\bar \varepsilon _{ps} ,\bar \mu _{ps} } \right],\\ 
 \bar \varepsilon _{ps}  = diag\left[ {\varepsilon _{r,p} ,\varepsilon _{\theta ,p} ,\varepsilon _{\phi ,p} } \right]\varepsilon _0 , \\ 
 \bar \mu _{ps}  = diag\left[ {\mu _{r,p} ,\mu _{\theta ,p} ,\mu _{\phi ,p} } \right]\mu _0 , \\ 
 \varepsilon _{r,p}  = \mu _{r,p}  = \frac{{R_3 }}{{R_3  - R_2 }}\frac{{(r - R_2 )^2 }}{{r^2 }}, \\ 
 \varepsilon _{\theta ,p}  = \varepsilon _{\phi ,p}  = \varepsilon _{\phi ,p}  = \mu _{\phi ,p}  = \frac{{R_3 }}
{{R_3  - R_2 }}, \\ 
 \end{array}
\end{equation}
as the  outer layer in (4) to construct a GLPS double layer cloak.
\subsection {A Single Negative Refraction Metamaterial}
For studying the EM wavefield excited by source inside a concealment which is cloaked by a
single layer cloak, we propose and situate a following  negative dielectric and positive susceptibility
metamaterial,
\begin{equation}
\begin{array}{l}
 \left[ D \right]_{GN}  = diag\left[ {\bar \varepsilon _{gn} ,\bar \mu _{gn} } \right], \\ 
 \bar \varepsilon _{gn}  = diag\left[ {\varepsilon _r ,\varepsilon _\theta  ,\varepsilon _\phi  } \right]\varepsilon _0 , \\ 
 \bar \mu _{gn}  = diag\left[ {\mu _r ,\mu _\theta  ,\mu _\phi  } \right]\mu _0 , \\ 
 \varepsilon _r  =  - \left( {\frac{{\sqrt {R_1^2  - r^2 } }}{r}} \right)^3 , \\ 
 \varepsilon _\theta   = \varepsilon _\phi   =  - \frac{r}{{\sqrt {R_1^2  - r^2 } }}, \\ 
\\
 \mu _r  = \left( {\frac{{\sqrt {R_1^2  - r^2 } }}{r}} \right)^3 , \\ 
 \mu _\theta   = \mu _\phi   = \frac{r}{{\sqrt {R_1^2  - r^2 } }}, \\ 
 \end{array}
\end{equation}
 inside the concealment, where $r \le R_1$ and $r \ge r_0 >  0$. Simulations by the GL EM modeling show that
the EM wavefield excited by sources inside the concealment with metamaterial $[D_{GN}]$ in (6) will propagate from the concealment to free space through
the single layer cloak. Therefore, the double layer cloak is necessary for sufficient invisiblity cloaking.

\section{3D GLOBAL AND LOCAL  EM MODELING}

In this section, 3D EM integral equations are presented briefly. Based on the integral
equations, we propose a 3D Global and Local EM field modeling.

\subsection{\label{sec:level1}3D ELECTROMAGNETIC INTEGRAL EQUATION}
The 3D EM integral equations in frequency domain have been  proposed in early papers
[1] and [2].  In this  paper,  we present the EM integral equations in time domain.
The integral equations can be derived from Maxwell's equations for the electric and magnetic fields, 
$E(r,t)$ and $H(r,t)$:

\begin{equation}
\begin{array}{l}
 \left[ {\begin{array}{*{20}c}
   {E(r,t)}  \\
   {H(r,t)}  \\
\end{array}} \right] = \left[ {\begin{array}{*{20}c}
   {E_b (r,t)}  \\
   {H_b (r,t)}  \\
\end{array}} \right] \\ 
  + \int\limits_\Omega  {G_{E,H}^{J,M} (r',r,t) * _t \delta \left[ {D(r')} \right]\left[ {\begin{array}{*{20}c}
   {E_b (r',t)}  \\
   {H_b (r',t)}  \\
\end{array}} \right]dr'}, \\ 
 \end{array}
\end{equation}
and
\begin{equation}
\begin{array}{l}
 \left[ {\begin{array}{*{20}c}
   {E(r,t)}  \\
   {H(r,t)}  \\
\end{array}} \right] = \left[ {\begin{array}{*{20}c}
   {E_b (r,t)}  \\
   {H_b (r,t)}  \\
\end{array}} \right] \\ 
  + \int\limits_\Omega  {G_{E,H,b}^{J,M} (r',r,t) * _t \delta \left[ {D(r')} \right]\left[ {\begin{array}{*{20}c}
   {E(r',t)}  \\
   {H(r',t)}  \\
\end{array}} \right]dr'}.  \\ 
 \end{array}
\end{equation}

In these equations, $E$ and $H$ are the total electric and magnetic fields in a given medium with prescribed EM sources; $E_b$ and $H_b$ are the 
fields in a background or reference medium with the same sources (the background fields are also called ``incident" fields); and 
$G_{E,H,b}^{J,M}$ in equation (8) is the EM Green's tensor for the background medium,
\begin{equation}
G_{E,H,b}^{J,M} \left( {r',r,t} \right) = \left[ {\begin{array}{*{20}c}
   {E_b^J \left( {r',r,t} \right)} & {E_b^M \left( {r',r,t} \right)}  \\
   {H_b^J \left( {r',r,t} \right)} & {H_b^M \left( {r',r,t} \right)}  \\
\end{array}} \right].
\end{equation}
The fields $E$, $H$, $E_b$, and $H_b$ are all vector quantities. 
The elements of the partitioned Green's tensor are 3$\times$3 tensors (matrices). For example, 
a point electric current source $J$ in the background medium produces the electric field $E_b^J$, while a point magnetic 
source $M$ produces the electric field $E_b^M$. Similar interpretations hold for $H_b^J$ and $H_b^M$. Also, 
$\delta \left[ D \right]$ is the electromagnetic material  parameter variation matrix,
\begin{equation}
\begin{array}{l}
 \delta \left[ D \right] = \left[ {\begin{array}{*{20}c}
   {\delta D_{11} } & 0  \\
   0 & {\delta D_{22} }  \\
\end{array}} \right], \\ 
 \delta D_{11}  = (\bar \sigma (r) - \sigma _b I) + (\bar \varepsilon (r) - \varepsilon _b I)\frac{\partial }{{\partial t}}, \\ 
 \delta D_{22}  = (\bar \mu (r) - \mu _b I)\frac{\partial }{{\partial t}}, \\ 
 \end{array}
\end{equation}
$\delta D_{11} $ and $\delta D_{22}  $ are a $3\times 3$ symmetry, 
inhomogeneous diagonal matrix  for  the isotropic material,
for anisotropic material, they are an inhomogeneous diagonal or full matrix, 
$I$ is a $3\times 3$ unit matrix; $\bar \sigma (r)$ is the conductivity tensor,
$\bar \varepsilon (r)$ is the dielectric tensor, $\bar \mu (r)$ is susceptibility tensor of the full medium, 
which can be dispersive (i.e., its properties can depend on the angular frequency $\omega$); 
$\sigma _b$ is the conductivity,  $\varepsilon_b $
is the permittivity, $ \mu _b$ is the permeability in the background medium; $\Omega$
is the finite domain in which the parameter variation matrix $\delta \left[ D \right] \ne 0;$
the $(\bar \varepsilon (r) - \varepsilon _b I)E $ 
is the induced electric polarization; and $(\bar \mu (r) - \mu _b I)H $ is the induced magnetization.
Finally, $ * _t$ is convolution with respect to time, $t$.

\subsection{\label{sec:level1}3D GL EM MODELING}
Although the two integral equations (7) and (8) are equivalent, there are numerical advantages in using both sets simultaneously.
Based on the integral equation,
the GL EM modeling in the time domain is proposed in this section. The 3D GL EM modeling and integral equations
in the frequency domain have been developed in early papers [1-2].

 (3.1)	The domain $\Omega$ is divided into a set of $N$ sub domains,$\{\Omega_k\}$, 
such that $\Omega  = \bigcup\limits_{k = 1}^N {\Omega _k }$. The division can be mesh or meshless.

(3.2)  When $k=0$, let
$E_0 (r,t)$ and $H_0 (r,t)$ are the analytical global field, $E_0^J (r',r,t)$, $H_0^J (r',r,t)$, $E_0^M (r',r,t)$, and
$H_0^M (r',r,t)$ are the analytical global Green's tensor in the background medium. By induction, suppose that
$E_{k-1} (r,t)$, $H_{k-1} (r,t)$, $E_{k-1}^J (r',r,t)$, $H_{k-1}^J (r',r,t)$, $E_{k-1}^M (r',r,t)$, and
$H_{k-1}^M (r',r,t)$ are calculated in the $(k-1)^{th}$ step in the subdomain $\Omega_{k-1}$.

(3.3) In $\{\Omega_k\}$, upon substituting $E_{k-1} (r,t)$, $H_{k-1} (r,t)$, $E_{k-1}^J (r',r,t)$, $H_{k-1}^J (r',r,t)$, $E_{k-1}^M (r',r,t)$, and
$H_{k-1}^M (r',r,t)$ into the integral equation (8),  the EM Green's tensor integral equation (8)
in $\Omega_{k}$ is reduced into $6\times 6$ matrix equations. By solving the $6\times 6$  matrix equations, 
we obtain the Green's tensor field  $E_{k}^J (r',r,t)$, $H_{k}^J (r',r,t)$, $E_{k}^M (r',r,t)$, and
$H_{k}^M (r',r,t)$.

(3.4) According to the  integral equation (7), the electromagnetic field $E_{k} (r,t)$ and $H_{k} (r,t)$ are updated by 
the interactive scattering field between the Green's tensor and local polarization and magnetization in the subdomain $\Omega_{k}$ as follows,
\begin{equation}
\begin{array}{l}
 \left[ {\begin{array}{*{20}c}
   {E_k (r,t)}  \\
   {H_k (r,t)}  \\
\end{array}} \right] = \left[ {\begin{array}{*{20}c}
   {E_{k - 1} (r,t)}  \\
   {H_{k - 1} (r,t)}  \\
\end{array}} \right] \\ 
  + \int\limits_{\Omega_k}  {\left\{ {\left[ {\begin{array}{*{20}c}
   {E_k^J (r',r,t)} & {H_k^J (r',r,t)}  \\
   {E_k^M (r',r,t)} & {H_k^M (r',r,t)}  \\
\end{array}} \right]} \right.}  \\ 
 \left. { * _t \delta \left[ {D(r')} \right]\left[ {\begin{array}{*{20}c}
   {E_{k - 1} (r',t)}  \\
   {H_{k - 1} (r',t)}  \\
\end{array}} \right]} \right\}dr' \\ 
 \end{array}
\end{equation}

(3.5) The steps (3.2) and (3.4) form a finite iteration, $k = 1,2, \cdots, N$,
the $E_N \left( r,t \right)$ and $H_N \left( r,t \right)$
are the electromagnetic field  of the GL modeling method. The GL
electromagnetic field modeling in the time space domain is short named as
GLT method.

The GL EM modeling in the space frequency domain is proposed in the paper \cite{2}, 
we call the GL modeling in frequency domain as GLF method. 

\section{\label{sec:level1} INTERACTION BETWEEN
THE EM WAVE FIELD WITH CLOAKS}

\subsection {Theoretical Analysis Of Interaction Between
The EM Wave Field With The GL Double Layer Cloaks}

 Theoretical analysis of the interaction between the EM wave with GL cloaks are presented in this section.

$\textbf{Statement 1:}$  Let domain $\Omega _{GL}$, equation (3), 
and anisotropic metamaterial $D_{GL}$, equation (4), be a ``GL" double layer cloak, 
and let the concealed central sphere $| \vec {r}  | < R_1$ and the region outside the cloak $ | \vec {r} | >R_3$
be filled with a normal electromagnetic material with permittivity and permeability,
 $\varepsilon {\rm  = }\varepsilon _{\rm b},\mu  = \mu _b$.  
We claim the following:
(1) provide EM wavefield is excited by sources located inside 
the concealment of GL cloak, $|\vec r_s |< R_1$, the EM wavefield 
never be disturbed 
by the cloak; (2) provide EM wavefield is excited by sources located inside 
concealment or the inner layer  of the cloak,
$|\vec r_s| < R_2$, the EM wavefield is absorbed in the inner layer $\Omega_{I}$ and does not extend outside the boundary $r=R_2$. 
(3) provide the wavefield is excited by  sources located outside 
of the cloak,  $|\vec r_s| >R_3,$ the  EM wavefield propagates as if in free space, undisturbed by the cloak; 
(4) provide the EM wavefield is excited by sources outside 
cloak or inside outer layer, $|\vec r_s| >R_2,$  
the EM wavefield does not penetrate into the inner layer and the concealed region, $|\vec r| <  R_2$. 

\subsection {No Maxwell EM Wavefield Can Be Excited By Nonzero
Local Sources Inside  The Single Layer Cloaked Concealment Fielled With Normal Materials}

$\textbf{Statement 2:}$ Suppose that a 3D anisotropic inhomogeneous single layer cloak domain separates the whole space into three sub domains: 
one is the single layer cloak domain $\Omega _{clk}$ with the cloaking material; another is the cloaked 
concealment domain $\Omega _{conl}$ with normal EM materials; and the third is 
free space outside the cloak. If the Maxwell EM wavefield excited 
by a point source or local sources outside the concealment $\Omega _{conl}$ vanishes
inside  the concealment $\Omega _{conl}$, then there is no Maxwell EM wave field excited 
by the local sources inside the cloaked concealment $\Omega _{conl}$.

The Maxwell EM wavefield is an  EM wave field that satisfies the Maxwell equation
and appropriate  interface boundary conditions. We use an inverse process to prove the $statement \ 2$ as follows:
Suppose that there exists a Maxwell EM wavefield excited by the local sources inside the 
concealment, filled with the normal materials, the wavefield will satisfy the Maxwell equation in all
of the space,
 including in the
anisotropic inhomogeneous cloaking layer $\Omega _{clk}$ and in the concealment, $\Omega _{conl}$; and also satisfies
boundary conditions on the interfaces, $S_1$ and $S_2$. 
$S_1$ is the interface boundary between the
cloaking region $\Omega _{clk}$ and the concealment domain $\Omega _{conl}$, and also is 
the inner boundary surface of the cloaking domain $\Omega _{clk}$.
$S_2$ is the interface boundary between the
cloaking region $\Omega _{clk}$ and free space outside; it also is 
the outer boundary surface of the cloaking region $\Omega _{clk}$.

Let $R_{c}  = R^3  - \Omega _{clk} \bigcup {\Omega _{conl}}$,  
$R_{d}  = R^3  - \Omega _{conl}$,
and by integral equation (7), the EM wave field satisfies
\begin{equation}
\begin{array}{l}
 \left[ {\begin{array}{*{20}c}
   {E\left( {r,t} \right)}  \\
   {H\left( {r,t} \right)}  \\
\end{array}} \right] = \left[ {\begin{array}{*{20}c}
   {E_b \left( {r,t} \right)}  \\
   {H_b \left( {r,t} \right)}  \\
\end{array}} \right] + \\ 
 \int\limits_{\Omega _{clk} \bigcup {\Omega _{con l} } } {G_{E,H}^{J,M} 
\left( {r',r,t} \right) * _t \delta \left[ D \right]\left[ {\begin{array}{*{20}c}
   {E_b \left( {r',t} \right)}  \\
   {H_b \left( {r',t} \right)}  \\
\end{array}} \right]dr'}, \\ 
 \end{array}
\end{equation}
where $G_{E,H}^{J,M} \left( {r',r,t} \right)$ is the  
Green's tensor, its components
$E^J$, $H^J$, $E^M$, and $H^M \left( {r',r,t} \right)$ are the
Green's function on $\Omega _{clk} \bigcup {\Omega _{conl} } \bigcup {R_c} $, 
excited by the 
point  impulse sources outside the concealment, $r \in R_d$.
By the assumptions, $G_{E,H}^{J,M} \left( {r',r,t} \right)$ exists on
$\Omega _{clk} \bigcup {\Omega _{conl} } \bigcup {R_c }$
and when $r' \in \Omega _{conl}$, $G_{E,H}^{J,M} \left( {r',r,t} \right) = 0$. 
The integral equation (12) becomes to
\begin{equation}
\begin{array}{l}
 \left[ {\begin{array}{*{20}c}
   {E\left( {r,t} \right)}  \\
   {H\left( {r,t} \right)}  \\
\end{array}} \right] = \left[ {\begin{array}{*{20}c}
   {E_b \left( {r,t} \right)}  \\
   {H_b \left( {r,t} \right)}  \\
\end{array}} \right] \\ 
  + \int\limits_{\Omega _{clk} } {G_{E,H}^{J,M} \left( {r',r,t} \right) * _t \delta \left[ D \right]
\left[ {\begin{array}{*{20}c}
   {E_b \left( {r',t} \right)}  \\
   {H_b \left( {r',t} \right)}  \\
\end{array}} \right]dr'.}  \\ 
 \end{array}
\end{equation}

Consider the Maxwell equation in $R_d$, the
virtual source is located $r$, $r \in R_d$ and the point source is located $r_s$, 
the variable in the following equations is $r'$, 
$ r_s  \in \Omega _{conl} $ 
and $ r_s  \notin R_d$, we have
\begin{equation}
\begin{array}{l}
 \left[ {\begin{array}{*{20}c}
   {} & {\nabla  \times }  \\
   { - \nabla  \times } & {}  \\
\end{array}} \right]G_{E,H}^{J,M} \left( {r',r,t} \right) \\ 
  = \left[ D \right]G_{E,H}^{J,M} \left( {r',r,t} \right) + I\delta (r',r)\delta (t), \\ 
 \end{array}
\end{equation}
and
\begin{equation}
\begin{array}{l}
 \left[ {\begin{array}{*{20}c}
   {} & {\nabla  \times }  \\
   { - \nabla  \times } & {}  \\
\end{array}} \right]\left[ {\begin{array}{*{20}c}
   {E_b }  \\
   {H_b }  \\
\end{array}} \right]\left( {r',r_s,t} \right) \\ 
  = \left[ {D_b } \right]\left[ {\begin{array}{*{20}c}
   {E_b }  \\
   {H_b }  \\
\end{array}} \right]\left( {r',r_s,t} \right), \\ 
 \end{array}
\end{equation}
Convolving equation (14) with $ \left[ {E_b \left( {r',t} \right),H_b \left( {r',t} \right)} \right]$, 
and equation (15) with $G_{E,H}^{J,M} \left( {r',r,t} \right)$, subtracting
the second result from the first, integrating by parts, and making some manipulations gives

\begin{equation}
\begin{array}{l}
 \left[ {\begin{array}{*{20}c}
   {E_b \left( {r,t} \right)}  \\
   {H_b \left( {r,t} \right)}  \\
\end{array}} \right] +  \\ 
  + \int\limits_{\Omega _{clk} } {G_{E,H}^{J,M} \left( {r',r,t} \right) * _t \delta \left[ D \right]\left[ {\begin{array}{*{20}c}
   {E_b \left( {r',t} \right)}  \\
   {H_b \left( {r',t} \right)}  \\
\end{array}} \right]dr'}  \\ 
  =  \oint\limits_{S_1 } {G_{E,H}^{J,M} \left( {r',r,t} \right) \otimes_t } \left[ {\begin{array}{*{20}c}
   {E_b \left( {r',t} \right)}  \\
   {H_b \left( {r',t} \right)}  \\
\end{array}} \right]d\vec S, \\ 
 \end{array}
\end{equation}

where $\otimes_t$ is cross convolution.  From the assumption of the $statement \ 2$ that "the Maxwell EM wavefield excited 
by a point source or local sources outside the concealment $\Omega _{conl}$
vanishes
inside the concealment $\Omega _{conl}$", and virtual source $r$ is located 
outside
the concealment, $r \in R_d$, if $ r' \in \Omega _{conl} $, 
$G_{E,H}^{J,M} \left( {r',r,t} \right) = 0$. By continuity,    
when $ r' \in S_1 $, we have 
$G_{E,H}^{J,M} \left( {r',r,t} \right) = 0$.  Continuity of
$G_{E,H}^{J,M} \left( {r',r,t} \right)$ at interface implies that the term in right hand side of (16) vanishes, giving
\begin{equation}
\begin{array}{l}
 \left[ {\begin{array}{*{20}c}
   {E_b \left( {r,t} \right)}  \\
   {H_b \left( {r,t} \right)}  \\
\end{array}} \right] +  \\ 
  + \int\limits_{\Omega _{clk} } {G_{E,H,b}^{J,M} \left( {r',r,t} \right) * _t \delta \left[ D \right]\left[ {\begin{array}{*{20}c}
   {E\left( {r',t} \right)}  \\
   {H\left( {r',t} \right)}  \\
\end{array}} \right]dr' = 0}. \\ 
 \end{array}
\end{equation}

Substituting equation (17) into the integral equation (13) gives
\begin{equation}
\left[ {\begin{array}{*{20}c}
   {E\left( {r,t} \right)}  \\
   {H\left( {r,t} \right)}  \\
\end{array}} \right] = 0.
\end{equation}

From continuity condition on the EM wave field, we obtain the following over vanishing  condition on the boundary $S_1$ of 
the concealment $\Omega _{conl}$,
\begin{equation}
\left. {\left[ {\begin{array}{*{20}c}
   {E\left( {r,r_s,t} \right)}  \\
   {H\left( {r,r_s,t} \right)}  \\
\end{array}} \right]} \right|_{S_1 }  = 0,
\end{equation}
where $r_s$ denotes point source location inside of the concealment, $r$ is EM field receiver point,
$r \in S_1$. 
Because the EM wavefield is excited by local sources inside of the concealment domain $\Omega _{conl}$, 
it satisfies
the following Maxwell equation with variable $r'$,
\begin{equation}
\begin{array}{l}
 \left[ {\begin{array}{*{20}c}
   {} & {\nabla  \times }  \\
   { - \nabla  \times } & {}  \\
\end{array}} \right]\left[ {\begin{array}{*{20}c}
   E  \\
   H  \\
\end{array}} \right]\left( {r',r_s,t} \right) \\ 
  = \left[ D_{conl} \right]\left[ {\begin{array}{*{20}c}
   E  \\
   H  \\
\end{array}} \right]\left( {r',r_s,t} \right) + Q(r',r_s,t). \\ 
 \end{array}
\end{equation}

where 
$\left[ D_{conl} \right] = diag\left[ {
   \varepsilon_r\varepsilon_0I,  \  \mu_r\mu_0 I} \right](\partial /\partial t)$
with the normal EM material parameters, $\varepsilon_r\ge 1$ and $\mu_r\ge 1$ are relative EM
parameters, $\varepsilon_0$ is basic permittivity and $\mu_0 $ is basic permeability, $I$ is $3 \times 3$
unit matrix,
$r_s  \in \Omega _{conl} $ is
the local source location, $Q(r',r_s,t)$ is the nozero local source inside the concealment, $ \Omega _{conl} $.
Let $G_{E,H,conl}^{J,M} (r',r,t)$ be $Green's$ tensor which satisfies 
\begin{equation}
\begin{array}{l}
 \left[ {\begin{array}{*{20}c}
   0 & {\nabla  \times }  \\
   { - \nabla  \times } & 0  \\
\end{array}} \right]G_{E,H,conl}^{J,M} (r',r,t) \\ 
  = \left[ {D_{conl} } \right]G_{E,H,conl}^{I,M} (r',r,t) \\ 
  + I\delta (r',r)\delta \left( t \right) \\ 
 \end{array}
\end{equation}

Convolving $ \left[ {E \left( {r',t} \right),H \left( {r',t} \right)} \right]$ with (21), 
and $G_{E,H,conl}^{J,M} \left( {r',r,t} \right)$ with (20), subtracting the second 
result from the first, taking the integral to $\Omega _{conl}$, and use integration by parts and other manipulations,
we have
\begin{equation}
\begin{array}{l}
 \left[ {\begin{array}{*{20}c}
   {E(r,r_s ,t)}  \\
   {H(r,r_s ,t)}  \\
\end{array}} \right] \\ 
  = \int\limits_{\Omega _{conl} } {G_{E,H,conl}^{J,M} } (r',r,t) * _t Q(r',r_s ,t)dr' \\ 
  + \oint\limits_{\partial \Omega _{conl} } {G_{E,H,conl}^{J,M} } (r',r,t) \otimes _t \left[ {\begin{array}{*{20}c}
   {E(r',r_s ,t)}  \\
   {H(r',r_s ,t)}  \\
\end{array}} \right]dr', \\ 
 \end{array}
\end{equation}
$\otimes _t $ denotes the cross convolution, and $\partial \Omega _{conl}=S_1$. Because of the over vanishing condition (19),
\[
\left. {\left[ {\begin{array}{*{20}c}
   {E\left( {r,r_s,t} \right)}  \\
   {H\left( {r,r_s,t} \right)}  \\
\end{array}} \right]} \right|_{S_1 }  = 0,			
\]
we have
\begin{equation}
\begin{array}{l}
 \left[ {\begin{array}{*{20}c}
   {E(r,r_s ,t)}  \\
   {H(r,r_s ,t)}  \\
\end{array}} \right] =  \\ 
  = \int\limits_{\Omega _{conl} } {G_{E,H,conl}^{J,M} } (r',r,t) * _t Q(r',r_s ,t)dr'. \\ 
 \end{array}
\end{equation}
Because ${G_{E,H,conl}^{J,M} } (r',r,t) \ne 0$ and $Q(r',r_s ,t) \ne 0$, so,
\begin{equation}
\left[ {\begin{array}{*{20}c}
   {E\left( {r,r_s,t} \right)}  \\
   {H\left( {r,r_s,t} \right)}  \\
\end{array}} \right] \ne 0.
\end{equation}
From the continuity of the EM wavefield, the nonzero EM wave field (24) results that 
\begin{equation}
\left. {\left[ {\begin{array}{*{20}c}
   {E\left( {r,r_s,t} \right)}  \\
   {H\left( {r,r_s,t} \right)}  \\
\end{array}} \right]} \right|_{S_1 }  \ne 0.
\end{equation}
The EM wavefield is nonzero on the boundasry $S_1$ in (25) that is an obvious contradiction with
the same EM wavefield is zero on the boundasry $S_1$ in (19).
Therefore, we proved statement 2.
Result (24) can also be derived more simply and obvious from the integral expression (23). Let the source be a point
impulse current source with polarization direction $\vec x$, i.e., 

\begin{equation}
Q(r,r_s ,t) = \delta \left( {r - r_s } \right)\delta (t)\vec x.
\end{equation}
Substituting (26) and $\varepsilon _r  =1.0 \ and\  \mu _r  = 1.0$ into the (23), gives
\begin{equation}
\left[ {\begin{array}{*{20}c}
   {E(r,r_s ,t)}  \\
   {H(r,r_s ,t)}  \\
\end{array}} \right] = \left[ {\begin{array}{*{20}c}
   {E_x^J (r,r_s ,t)}  \\
   {H_x^J (r,r_s ,t)}  \\
\end{array}} \right]
\end{equation}
\begin{equation}
E_x^J (r,r_s ,t) = \left[ {\begin{array}{*{20}c}
   {E_{xx} (r,r_s ,t)}  \\
   {E_{xy} (r,r_s ,t)}  \\
   {E_{xz} (r,r_s ,t)}  \\
\end{array}} \right]
\end{equation}
\begin{equation}
\begin{array}{l}
 E_{xx} (r,r_s ,t) \\ 
  =  - \frac{1}{{8\pi ^2 \varepsilon }}\frac{{\partial ^2 }}{{\partial x^2 }}\frac{{\delta \left( {t - \sqrt {\varepsilon \mu } \left| {r - r_s } \right|} \right)}}{{\left| {r - r_s } \right|}} \\ 
  + \frac{1}{{8\pi ^2 }}\mu \frac{{\partial ^2 }}{{\partial t^2 }}\frac{{\delta \left( {t - \sqrt {\varepsilon \mu } \left| {r - r_s } \right|} \right)}}{{\left| {r - r_s } \right|}} \\ 
 \end{array}
\end{equation}
It is obvious that when $r \in S_1$
\begin{equation}
\left. {E_{xx} (r,r_s ,t)} \right|_{r \in S_1 }  \ne 0.
\end{equation}
The electric intensity field  $\left. {E_{xx} (r,r_s ,t)} \right|_{r \in S_1 }  \ne 0$ in (30)
and $\left. {E_{xx} (r,r_s ,t)} \right|_{r \in S_1 }  = 0.$ in (19)
are an obvious contradiction. The $Statement \ 2$ is true.

\section{\label{sec:level1} Simulations of The EM Wave Field 
Through The Double Layer Cloak}
\subsection{The Model of The Double Layer Cloak}
The full 3D simulation model is a unit cube, $[0.5, 0.5]^3$, centered on the
origin, discretized on a $201^3 $ mesh, with uniform mesh spacing of $0.005m$.
The EM wavefield is excited by a point electric source:
\begin{equation}
s(r,r_s ,t) = \delta (r - r_s )\delta (t)\vec e,
\end{equation}
at location $r_s$ where $\vec e$  is the (unit) polarization vector. the time
step $dt = 0.3333 \times 10^{ - 10}$ ; the frequency band is from $0.05GHz$ to $15GHz$.
The shortest wavelength is about $0.02m$. 
The GL double layer EM cloak consists of inner and outer annular regions,
$\Omega _{GL}  = \Omega _{I} \bigcup {\Omega _{O} }$, equation (3), 
with centers at the origin,  which is situated by proposed
anisotropic metamaterial $D_{GL}$, equation (4), the concealed central sphere $| \vec {r}  | < R_1$ 
and the region outside the cloak $ | \vec {r} | >R_3$
is filled with a normal electromagnetic material with basic permittivity and permeability,
 $\varepsilon {\rm  = }\varepsilon _{\rm b},\mu  = \mu _b$.  
The inner boundary of the cloak is
$R_1= 0.2m$; the middle shell boundary between the two layers is $R_2= 0.3m$; and the
outer boundary is $R_3= 0.45m.$
In the simulations, the sphere region $r \le R_3 $  is
actually modeled in spherical coordinates,   $ \  (r,\theta ,\phi )$, where $\theta $
is polar angle. The sphere is divided $180^3$ cells. 
The spherical coordinate grid is
superimposed on the rectangular 
grid used to mesh the domain outside shell $r=R_3$.
Because the cloaking materials are radial dependent,
the  EM modeling is reduced to the system of the one dimensional GL modeling by using the sphere harmonic expansion.

\begin{figure}[h]
\centerline{\includegraphics[width=0.86\linewidth,draft=false]{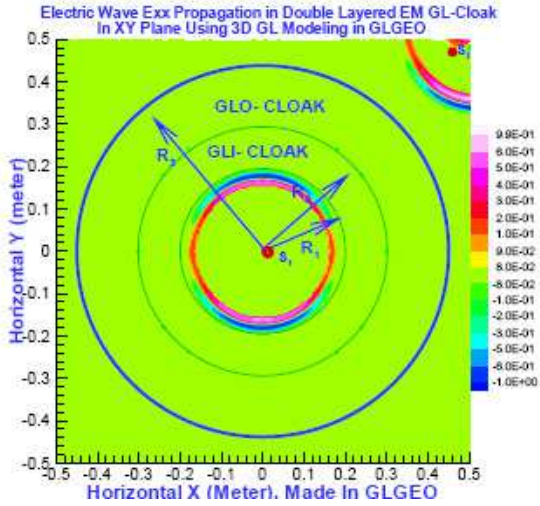}}
\caption{(color online) At time step $18dt$, the wave front of the $ {\it First \  electric \  wave} $, $E_{xx,1}$, propagates inside the 
concealment $r < R_1$. The front of $ {\it Second \  EM\  wave} $, $E_{xx,2}$, 
is located in free space,  the right and top corner outside of the whole GL double layer cloak.}  \label{fig1}
\end{figure}

\begin{figure}[h]
\centerline{\includegraphics[width=0.86\linewidth,draft=false]{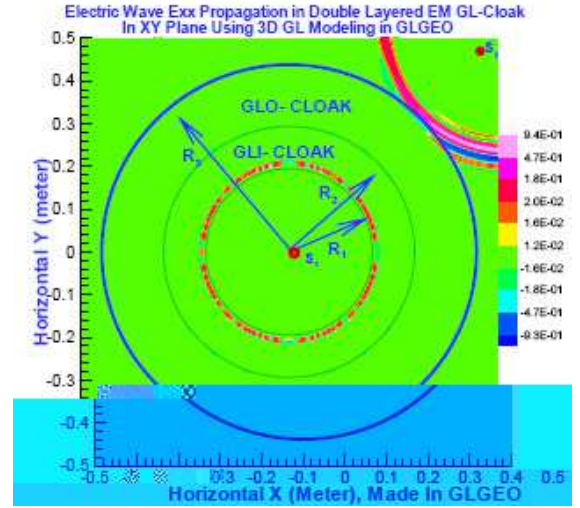}}
\caption{ (color online)  At this moment of the time step 30dt, the front of the $ {\it First \  electric\  wave} $,  $E_{xx,1}$,
 propagates enter to the inner layer, $R_1 \le r \le R_2$;  the front of $ {\it Second \  electric\  wave} $,  
$E_{xx,2}$,  reaches the outer boundary $r=R_3$ of the
GL double layer cloak.}\label{fig2}
\end{figure}

\begin{figure}[h]
\centerline{\includegraphics[width=0.86\linewidth,draft=false]{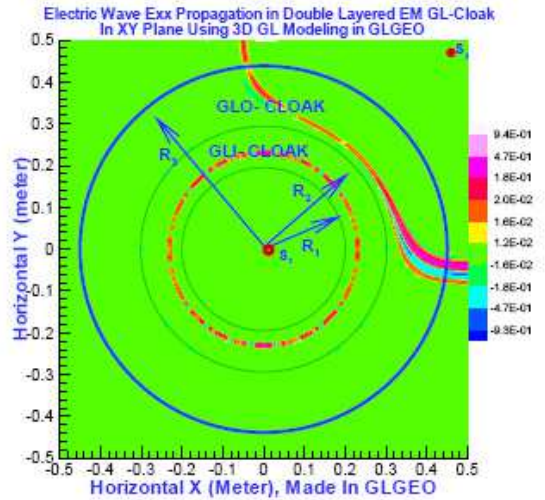}}
\caption{ (color online)  At the time step $58dt$, the ${\it First \  electric\  wave} $, $E_{xx,1}$, is propagating  inside 
the inner layer, $R_1 \le r \le R_2$, and becomes very slow. The part of the front of ${\it Second \  electric\  wave} $,  $E_{xx,2}$,  has been inside of
outer layer, $R_2 \le r \le R_3$, and being backward bending }\label{fig3}
\end{figure}

\begin{figure}[h]
\centerline{\includegraphics[width=0.86\linewidth,draft=false]{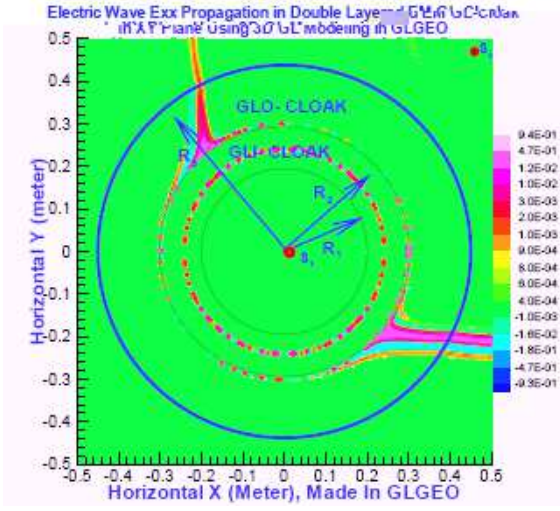}}
\caption{ (color online) At time step $75dt$, the $ {\it Second \  electric\  wave}$, $E_{xx,2}$,
is propagating inside of the outer layer, 
$R_2 \le r \le R_3$, and around the shell $r=R_2$ and never penetrate into inner layer and concealment,
${r < R_2}$. It does split into the two phases around the shell $r=R_2$. 
The $ {\it First \  electric\  wave} $,  $E_{xx,1}$, is propagating inside 
 the inner layer, $R_1 \le r \le R_2$.}\label{fig4}
\end{figure}
\begin{figure}[h]
\centerline{\includegraphics[width=0.86\linewidth,draft=false]{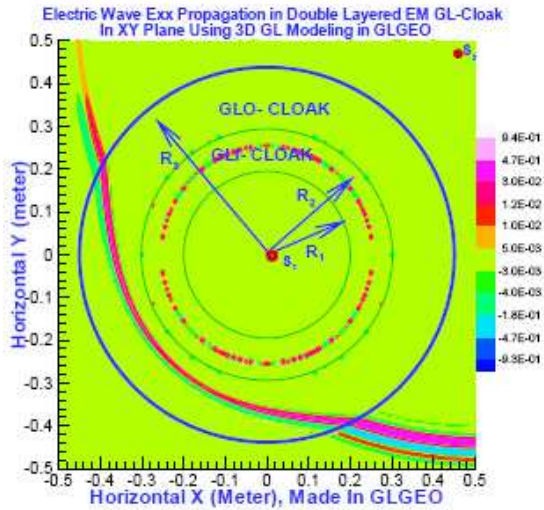}}
\caption{ (color online)  At time step $98dt$, one part of front 
the $ {\it Second \  electric\  wave}$, $E_{xx,2}$,  is propagating inside of the outer layer, $R_2 \le r \le R_3$, and
around to other side far source. It is a little
forward bending. The $ {\it First \  electric\  wave} $,  $E_{xx,1}$,   is still propagating inside 
the inner  layer, $R_1 \le r \le R_2$.}\label{fig5}
\end{figure}
\begin{figure}[h]
\centerline{\includegraphics[width=0.86\linewidth,draft=false]{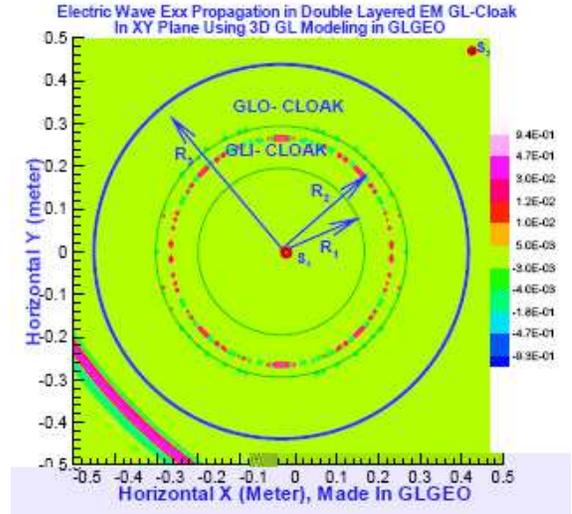}}
\caption{ (color online)  At time step $128dt$, 
the $ {\it Second \  electric \  wave}$,  $E_{xx,2}$
has propagated outside the GL double layer cloak, a small part of its wave front is located in the left and low corner of the plot frame,
which never be disturbed. 
The $ {\it First \  electric \  wave}$,  $E_{xx,1}$, 
is still propagating inside the inner layer, $R_1 \le r \le R_2$.}\label{fig6}
\end{figure}

\begin{figure}[h]
\centerline{\includegraphics[width=0.86\linewidth,draft=false]{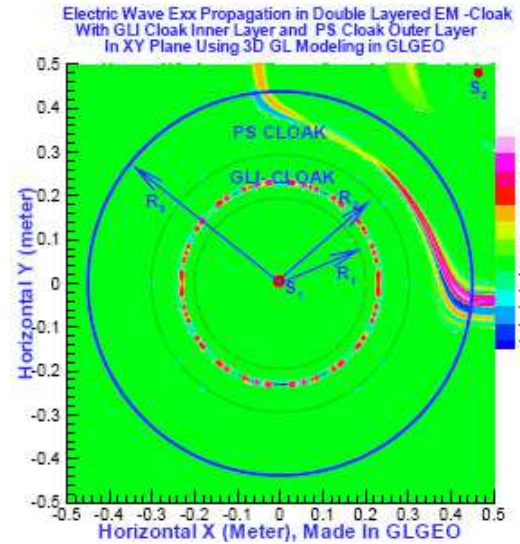}}
\caption{ (color online) In the double layer cloak, inner layer is situated with GLI cloak material in (1),
outer layer is filled with PS cloak material.
At the time step $58dt$, the 
${\it First \  electric\  wave} $, 
$E_{xx,1}$, is propagating inside inner layer; The $ {\it Second \  electric\  wave}$ propagates
inside outer layer with more backward bending and more slow than EM wavefield in GLO material (2)
in Figure 5.}\label{fig7}
\end{figure}

\begin{figure}[h]
\centerline{\includegraphics[width=0.86\linewidth,draft=false]{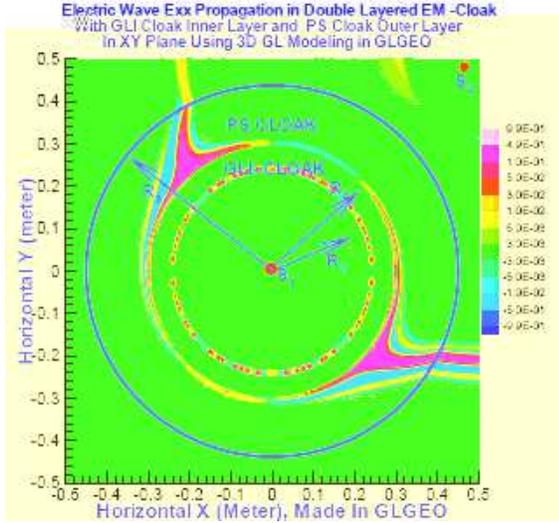}}
\caption{ (color online) In the double layer cloak GLPS arranged in Figure 7, at the time step $75dt$, the 
${\it First \  electric\  wave} $, 
$E_{xx,1}$, is propagating inside inner layer; The $ {\it Second \  electric\  wave}$ propagates
around the shell $r=R_2$ with more forward bending  than EM wavefield in GLO material (2)
in Figure 5.}\label{fig8}
\end{figure}

\begin{figure}[h]
\centerline{\includegraphics[width=0.86\linewidth,draft=false]{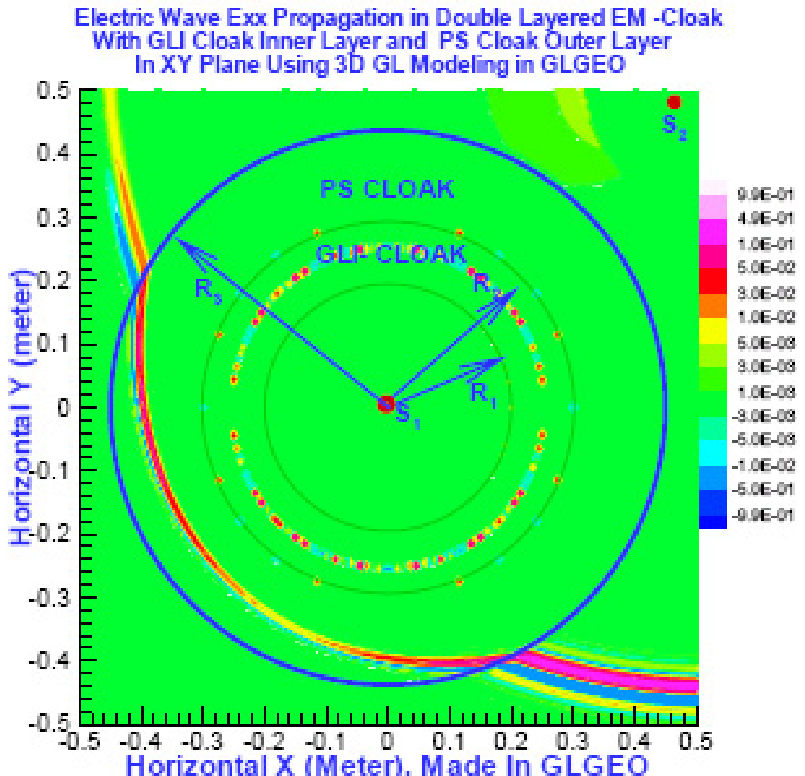}}
\caption{ (color online) In the double layer cloak GLPS in Figure 7,  at the time step $98dt$, the 
${\it First \  electric\  wave} $, 
$E_{xx,1}$, is propagating inside inner layer; The $ {\it Second \  electric\  wave}$ propagates inside outer layer
around to side far source with more forward bending and more fast than EM wavefield does in GLO material (2)
in Figure 5.}\label{fig9}
\end{figure}

\subsection{Simulation I}

In this subsection, the simulation I is presented that
an inner point source in the concealment and other outer source in the free space are used to excite
the EM wave propagation through the  double layer cloak.
The point
impulse current source with polarization direction $\vec x$, i.e.

\begin{equation}
S(r,r_s ,t) = \delta \left( {r - r_s } \right)\delta (t)\vec x.
\end{equation}

The first point current source is located inside the  concealment 
at $(0.0012m,0.0m,0.0m)$,
by which the
excited EM wave is named as $ {\it First \  EM\  wave} $,  its component $E_{xx,1}$ is labeled
$ {\it First \  electric \  wave} $,  
The second current point source is located in free space at $(0.518m,0.518m,0.0)$ where 
is in the right and top corner outside the  double layer cloak. 
The EM wave by the second source is named as $ {\it Second \  EM\  wave} $.
Its component $E_{xx,2}$, is labeled $ {\it Second \  electric \  wave} $.
Figure 1-6 show a series of snapshots of an EM field propagating in and
around the double layer cloak. 
The Figure 1 shows that at time step $18dt$, the wave front of the $ {\it First \  electric \  wave} $,  
$E_{xx,1}$, propagates inside the central sphere concealment $r < R_1$ and never
be disturbed by the inner layer;
The front of $ {\it Second \  EM\  wave} $, $E_{xx,2}$, 
is located in free space, the right and top corner outside the  double layer cloak. 
At the moment $30dt$ in Figure 2, the front of the 
$ {\it First \  electric\  wave} $,  $E_{xx,1}$, propagates enter to the 
inner layer, $R_1 \le r \le R_2$; The front of
$ {\it Second \  electric\  wave} $,  $E_{xx,2}$,  reaches the outer shell boundary $r=R_3$ of the
 cloak and never be disturbed. At the time step $58dt$, the $ {\it First \  electric\  wave} $, 
 $E_{xx,1}$,  is  propagating inside the inner layer, $R_1 \le r \le R_2$, and becomes very slow; 
The part of the front of
$ {\it Second \  electric\  wave} $,  $E_{xx,2}$,  propagates inside outer layer, 
$R_2 \le r \le R_3$, and being backward bending. 
The EM wave propagation image snapshot is presented in Figure 3.  In the Figure 4, at time step $75dt$, 
the $ {\it First \  electric\  wave} $, 
 $E_{xx,1}$,  is still propagating inside the inner layer, $R_1 \le r \le R_2$;
The $ {\it Second \  electric\  wave}$, $E_{xx,2}$,  is propagating inside the outer layer cloak,
$R_2 \le r \le R_3$,
and around the shell $r=R_2$ and never penetrate into the inner layer  and concealment, $r \le R_2$. 
It does split into the two phases around the shell $r=R_2$, 
the front phase speed exceeds the light speed; the back phase is slower than the light speed. 
In the figure 5, at time step $98dt$, one part of front 
the $ {\it Second \  electric\  wave}$, $E_{xx,2}$,  is propagating inside  the outer layer 
, $R_2 \le r \le R_3$. has a little forward bending
and never penetrate into the inner layer and the concealment, i.e. $r \le R_2$. 
The $ {\it First \  electric\  wave} $,  $E_{xx,1}$,   is still propagating inside 
of the inner layer, $R_1 \le r \le R_2$.
At time step $128dt$, 
the $ {\it Second \  electric \  wave}$,  $E_{xx,2}$
has propagated outside  double layer cloak, a small part of its wave front is 
located in the left and
low corner of the plot frame which is shown in the
Figure 6, most part of front of the $E_{xx,2}$ electric wave field has been out
of the plot frame  The exterior EM wave outside  the cloak 
never been disturbed 
by the cloak and never penetrate enter
its concealment and inner layer. 
At same time step, the $ {\it First \  electric \  wave}$,  $E_{xx,1}$, 
is still propagating inside the 
the inner layer, $R_1 \le r \le R_2$. It can be very 
closed to the interface boundary shell $r = R_2$, However, it can not be reached to 
the boundary $r = R_2 $ for any long time. That means that the interior EM wavefield
excited by source inside the concealment is complete absorbed by the inner layer.
The inner layer metamaterial, in equation (1),  cloaks outer space from the local field excited in the inner layer
and concealment,  which can also be useful for making a complete absorption boundary condition
to truncate infinite domain in numerical simulation. 

\begin{figure}[h]
\centerline{\includegraphics[width=0.86\linewidth,draft=false]{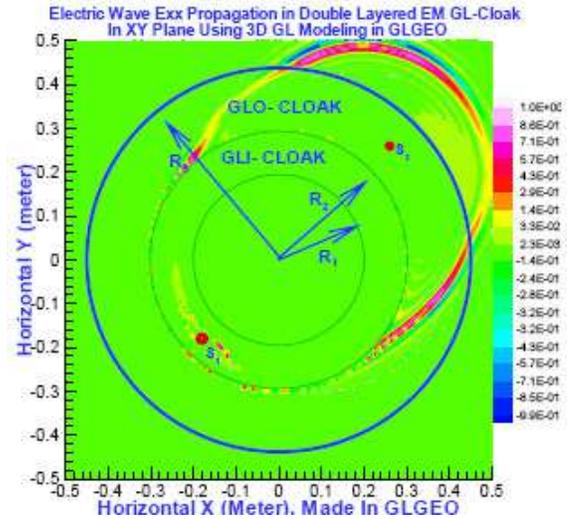}}
\caption{ (color online)  At time step $30dt$,  one part of the front of 
the $ {\it Second \  electric \  wavefield}$,  $E_{xx,2}$,
propagates in free space with disturbance,  its other part of the front
has been propagating around the shell $r=R_2$;  It does not propagate
into the inner layer and concealment. The $ {\it First \  electric \  wave}$,  $E_{xx,1}$, 
is very slow and just starts to propagate in the inner layer.}\label{fig10}
\end{figure}

\begin{figure}[h]
\centerline{\includegraphics[width=0.86\linewidth,draft=false]{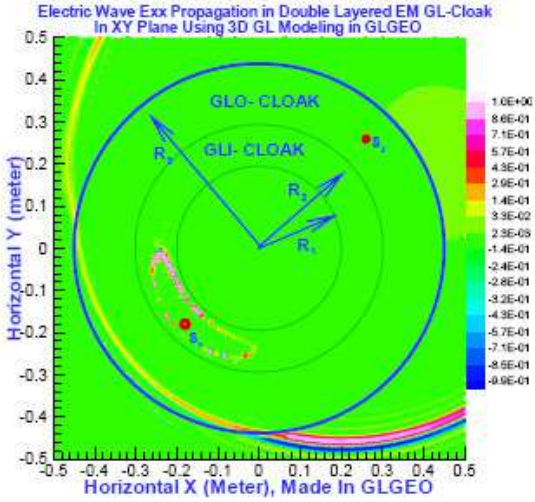}}
\caption{ (color online) 
At time step $68dt$,  front of the $ {\it First \  electric \  wave} $,  
$E_{xx,1}$, is very slow propagating inside inner layer $R_1< r < R_2$; The most part of front of $ {\it Second \  EM\  wave} $, $E_{xx,2}$, 
propagates in free space with disturbance, its other small part of front propagates in the
outer layer around to side far source, but it does not propagate into the inner layer. }\label{fig11}
\end{figure}

\begin{figure}[h]
\centerline{\includegraphics[width=0.86\linewidth,draft=false]{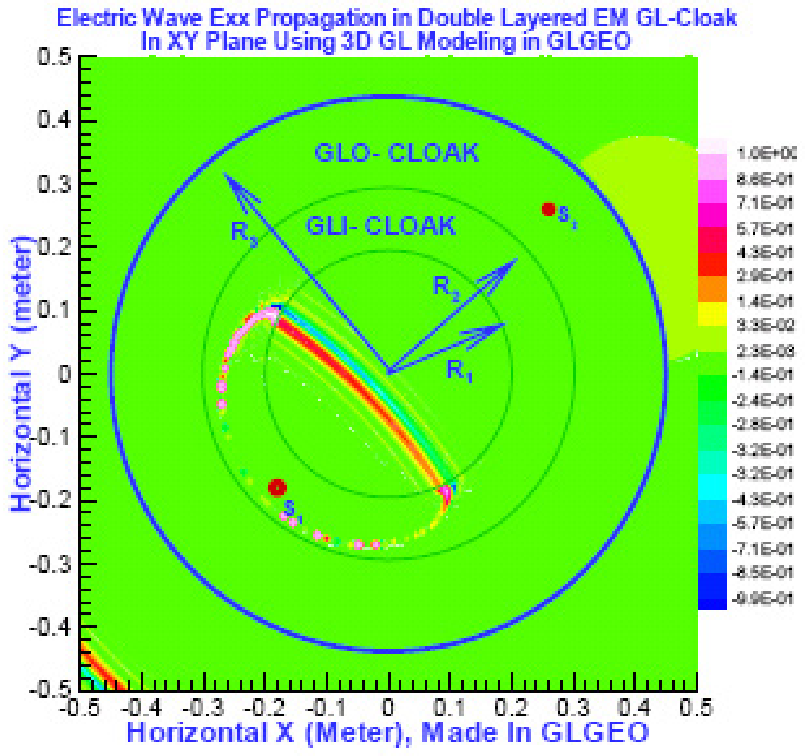}}
\caption{ (color online)
At moment $98dt$,  
one part of the front of the 
$ {\it First \  electric\  wave} $,  $E_{xx,1}$, propagates enter to the 
concealment; other part of the front is still propagating inside of
inner  layer, $R_1 \le r \le R_2$., it does not propagate outside shell $r=R_2$;   The front of
$ {\it Second \  electric\  wave} $,  $E_{xx,2}$,  has propagated outside all cloak, only very small
part of the front is located the left lower corner of the  figure frame.}
\label{fig12}
\end{figure}

\begin{figure}[h]
\centerline{\includegraphics[width=0.86\linewidth,draft=false]{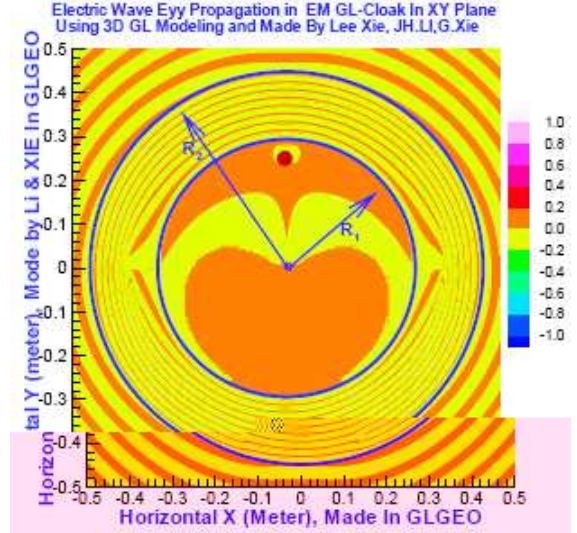}}
\caption{ (color online) 
The PS single layer cloaked concealment is filled with a permittivity single negative 
metamaterial, The electric wavefield Exx excited by a point source inside the concealment
propagates out to free space through the cloak.
}\label{fig13}
\end{figure}

\subsection{Simulation II}

To compare the properties between the GLO cloak and PS cloak, we use PS cloak [4] as outer layer and GLI cloak,
in equation (1),  as inner layer to construct a GLPS double layer cloak. The simulations of the EM wavefield propagation
through the GLPS cloak is presented in simulation II. The cloak geometry and mesh configuration is described in
subsection A.  Because the GLI cloak, in equation (1),  is used as inner cloak of GL double layer cloak and GLPS
double layer cloak, the  $ {\it First \  electric \  wave} $
propagation in timestep 30dt, 58dt, and 98dt are as  same as  in Figure 3 and Figure  7; Figure 4 and Figure 8, Figure 5
and Figure 9, respectively. At timesep 58dt, the $ {\it Second \  electric\  wave}$, $E_{xx,2}$, propagating in 
PS cloak is more slow and more backward bending than the field does in GLO outer layer, in Figure 3.
At timestep 75dt, the front of $ {\it Second \  electric\  wave}$, $E_{xx,2}$, propagating in  outer layer of GL cloak in Figure 4,  
is better than the field propagation in PS cloak, in Figure 8.
In particular, at timestep 98dt,  $ {\it Second \  electric\  wave}$, $E_{xx,2}$, propagating in 
layer PS cloak has more foeward bending in Figure 9, but the field has a little forward bending in GLO outer
layer in figure 4. Summary, GLO cloak material degeneration is weaker and better than the PS cloak.

\subsection{Simulation III}

The EM wavefield excited by a point source inside the inner layer and other point source
inside  the outer layer of GL cloak propagates through the double layer cloak
that is presented in this subsection. By the first point current source at $(-0.18m,-0.18m,0.0)$, the
excited EM wavefield $E_{xx,1}$ is labeled $ {\it First \  electric \  wave} $.  
The second current point source at $(0.258m,0.258m,0.0)$ excites 
the EM wavefield, $E_{xx,2}$, which is labeled $ {\it Second \  electric \  wave} $.
The Figure 10 shows that at time step $30dt$,  one part of the front of 
the $ {\it Second \  electric \  wavefield}$,  $E_{xx,2}$,
propagates in free space with disturbance in [15],  its other part of the front
has been propagating around the shell $r=R_2$;  It does not propagate
into the inner layer and concealment. The $ {\it First \  electric \  wave}$,  $E_{xx,1}$, 
is very slow and just starts to propagate in the inner layer.
In the Figure 11, at time step $68dt$,  front of the $ {\it First \  electric \  wave} $,  
$E_{xx,1}$, is very slow propagating inside inner layer $R_1< r < R_2$; The most part of front of $ {\it Second \  EM\  wave} $, $E_{xx,2}$, 
propagates in free space with disturbance, its other part of front propagates inside the
outer layer around to side far source, but it does not propagate into the inner layer.
The wavefield propagating at moment $98dt$ is presented in figure 12. 
One part of the front of the 
$ {\it First \  electric\  wave} $,  $E_{xx,1}$, propagates enter to the 
concealment; other part of the front is still propagating inside of
inner  layer, $R_1 \le r \le R_2$., it does not propagate outside shell $r=R_2$;   The front of
$ {\it Second \  electric\  wave} $,  $E_{xx,2}$,  has propagated outside all cloak, only very small
part of the front is located the left lower corner of the  figure frame
\\ \\  \\

\subsection{Simulation IV}

For studying the EM wavefield excited by source inside a concealment which is cloaked by a
single layer cloak, we propose and situate a novel  negative dielectric and positive susceptibility
metamaterial $[D]_{GN}$, in equation (6),
inside the concealment, where $r \le R_1$ and $r \ge r_0 >  0$. The concealment is cloaked by the PS cloak [4],
in equation (5). The simulation domain and mesh configration is described in subsection A.
A time harmanic point source with polarization in y direction and frequency  $0.57 \times 10^{10} Hz$,
\begin{equation}
s(r,r_s,t) = \delta (r - r_s )e^{i\omega t} \vec y,
\end{equation}
is located in the  point $r_s$, $(0.0,0.26,0.0)$.
 Simulations by the GL EM modeling show that, in Figure 13,
the EM wavefield  $E_{yy}$ excited by the source in (33) inside the concealment filled with the  metamaterial is propagating from the concealment 
to free space through
the PS single layer cloak. Therefore, the double layer cloak is necessary for sufficient invisiblity cloaking. This single negative refraction metamaterial,
in equation (6), will be investigated and detailed presented in next paper.

\section{\label{sec:level1}Advantages}

\subsection {The EM GL Double Layer Cloak Is Robust and Sufficient  For Invisibility}
The figure 10-12 clearly show that at moment of the 30dt, 68dt, and 98dt time step, the 
wave front of the Second electric wave has propagated outside the 
cloak and go to free space with disturbance. The result reminders us that 
if only single outer layer cloak $\Omega_{O}$, or PS cloak is adopted, and there is 
a little crack loss on the inner side of the boundary surface $\partial{{\Omega_{O}}_-}$, 
some EM or current source inside the $\Omega_{O}$
will excite the EM wave propagation go out to free space and expose the 
cloak immediately.  Moreover, the Figure 13 shows that the EM wavefield  $E_{yy}$ excited by the 
source in (33) inside the concealment filled with the  metamaterial is propagating from the concealment to free space through
the PS single layer cloak. Such that the single layer cloak
complete lose  the cloaking function.  The GL double layer cloak overcomes the weakness that
is also shown in the figure 1-12.
The wave front of the First electric wave, which is excited by a point
source inside the inner layer cloak $\Omega_{I}$, or inside the concealment, is always propagating 
inside of inner layer cloak $\Omega_{I}$ or
concealment $\Omega_{conl}$ and never propagate outside of the interface 
shell $ r=R_2 $. Any EM field inside the inner layer or inside the concealment can not
be propagated outside the shell $r=R_2$. Therefore, the EM GL double layer cloak is robust
and sufficient for invisibility.

Using the GL method theoretical analysis, the statement 2 in section 4 is rigorously 
proved. It states that "there exists no Maxwell electromagnetic wavefield can be
excited by nonzero local sources inside of the single layer cloaked concealment
with the normal EM materials". The invisibility of the single layer cloak
and existence of Maxwell EM wave field excited by the local sources inside 
its concealment is inconsistent. Provide only single outer layer cloak
is adopted, the EM field excited by local sources inside of its concealment with
normal materials does not satisfy the Maxwell equation. The EM chaos phenomena, 
which is divorced from the Maxwell equation governing, may damage devices
and human inside the concealment, or may degrade the invisibility of the cloak.
The single layer cloak is not complete and unsafe.
The Figure 13 shows when a special metamaterial fills into the concealment, the single layer cloak complete lose cloaking function.
The GL  double layer cloak in this paper or double coating in paper [8] are necessary for
the complete invisibility function. The GL double layer cloak is different from the double coating in paper [8].  
The simulations in Figures 1-12 show that the inner layer metamaterial of GL double layer,
in equation (1), does not disturb and reflect the EM field excited by sources inside the concealment
and does not change the EM environment in the concealment. Therefore, GL cloak
is safe metamaterial.

\subsection {Freqency Band And Geometry}

Many simulations and theoretical analysis by the GL method show that the idea
EM GL double layer cloak is of the invisibility function for all frequencies. 
However, the practical material has some loss.
The frequency band will be depended on the rate of the material loss. 

When $r$ is decreasing and going to $R_2$ in outer layer, the EM wavefield speed in PS cloak,
In equation (5), is increasing to infinite with the strong divergent rate $1/(r-R_2)$, but the EM
wavefield speed in outer layer of GL cloak, in equation (1), is increasing with the weak 
divergent rate $1/\sqrt{r-R_2}$. In the side near the source, the simulation in 
Figure 7 shows that $ {\it Second \  electric\  wave} $, $E_{xx,2}$, propagation is 
more slow and more backward bending inside the PS cloak than the field does in 
outer layer of GL cloak in Figure 3. In the side far the source, the simulation in 
Figure 9 shows that $ {\it Second \  electric\  wave} $,  $E_{xx,2}$,   propagation 
is more fast and more forward bending inside the PS cloak than the field does in outer layer of GL cloak in Figure 5.

The comparisons between the EM wave propagation inside the GL cloak and PS cloak,
in Figure 3 - 5 and Figure 7 - 9,  show that the GL cloak has better properties to
reduce the dispersion and degeneration.  Therefore, 
a reasonable  frequency band of the GL double layer cloak for low loss rate 
may be obtained in practical fabrication. The EM GL double layer cloak can be extended to have double ellipsoid annular layer
and other double strip geometry or use non edclidean geometry [10]
in outer layer cloak and GLI in (1) in inner layer to make wide frequency band double layer cloak in next paper. The experiments in this field, the pioneering
demonstration by D. Schurig et al in [11] and the recent
paper by R. Liu et al. in [12] may be important  help for improving
cloak model and simulation. As early in 2001, a double layer cloth phenomenon has been 
observed in Lawrence Berkely National Laboratory which is published in SEG  Expanded Abstracts
in 2002 [13].
For studying and simulating the strange phenomenon and metamaterials, 
we developed the effective GL modeling 
and inversion. The first paper of the GL method and a relative EM ``mirage" phenomenon
have been presented in PIERS 2005 and published in the 
proceeding of PIERS 2005 in Hangzhou in [14] and [15].

\subsection {Advantages Of The GL Method}

The GL EM modeling is fully different from FEM and FD and Born approximation methods and overcome their difficulties. There is no big matrix equation to solve in GL method.
Moreover, it does not need artificial boundary and absorption condition
to truncate the infinite domain. 
Born Approximation is a conventional method in the quantum mechanics
and solid physics, however, it is one iteration only in whole domain which is
not accurate for high frequency and for high contrast materials. The GL method divides the domain as a set of small 
sub domains or sub lattices. The Global field is updated by the local field 
from the interaction between the global field  
and local subdomain materials successively. Once all subdomain
materials are scattered, the GL field solution is obtained which
is much more accurate than the Born approximation. GL method
is suitable for all frequency and high contrast materials.
 
Moreover, the GL method can be meshless, including arbitrary geometry
subdomains, such as rectangle, cylindrical and spherical coordinate
mixed coupled together. It is full parallel algorithm.  
The GL EM method consistent combines the 
analytical and numerical approaches together and reduced the numerical 
dispersion and numerical frequency limitation.
The GL method has double capabilities of
the theoretical analysis and numerical simulations that
has been shown in this paper.
Because the cloak metamaterial in (4), (5),
and (6) are radial dependent, the reduced one dimension GL EM modeling system by using the  sphere
harmonic expansion are developed for the cloak simulations.
The 3D GL simulations of the EM wave field through the single and multiple sphere, cylinder, ellipsoid, and arbitrary geometry cloaks in single layer and double layer cloaks
show that the GLT and GLF EM modeling are accurate, stable and fast.
 The 3D and 2D GL parallel software are  developed and patented by GLGEO.

\section{\label{sec:level1}CONCLUSIONS}

Many simulations by the GL modeling and theoretical analysis verify that the EM GL doubled
cloak is robust and safe cloak and has complete and sufficient
invisibility functions. Its concealment is the normal electromagnetic
environment.
The outer layer of the GL double layer cloak has the invisible function,
its inner layer cloak has fully absorption function.
The GL method is an effective physical simulation method.
It has double capability of the theoretical analysis and numerical simulations to study the cloak metamaterials and wide material and 
Field scattering in physical sciences.

\begin{acknowledgments}
We wish to acknowledge the support of the GL Geophysical Laboratory.
Authors thank to Professor P. D. Lax for his concern and encouragements.
\end{acknowledgments}


\end{document}